\documentclass[manuscript, nonacm]{acmart}
\AtBeginDocument{%
  }

\setcopyright{none}                 
\begin{document}

\title{Designing and Evaluating Malinowski's Lens: An AI-Native Educational Game for Ethnographic Learning}

\author{Michael Hoffmann}

\affiliation{%
  \institution{Freie Universität Berlin}
  \city{Berlin}
  \country{Germany}
}
\email{michaeh78@zedat.fu-berlin.de}

\author{Jophin John}
\affiliation{%
  \institution{Leibniz Supercomputing Centre}
  \city{Garching near Munich}
  \country{Germany}}
\email{Jophin.John@lrz.de}

\author{Jan Fillies}
\affiliation{%
  \institution{Stanford University}
  \city{Stanford}
  \country{USA}}
\email{fillies@stanford.edu}
\affiliation{%
  \institution{Freie Universität Berlin}
  \city{Berlin}
  \country{Germany}}
\email{fillies@zedat.fu-berlin.de}

\author{Adrian Paschke}
\affiliation{%
  \institution{Fraunhofer Fokus}
  \city{Berlin}
  \country{Germany}
}
\email{adrian.paschke@fokus.fraunhofer.de}
\affiliation{%
  \institution{Freie Universität Berlin}
  \city{Berlin}
  \country{Germany}}
\email{adrian.paschke@fu-berlin.de}

\begin{abstract}
This study introduces 'Malinowski's Lens', the first AI-native educational game for anthropology that transforms Bronisław Malinowski's 'Argonauts of the Western Pacific' (1922) into an interactive learning experience. The system combines Retrieval-Augmented Generation with DALL·E 3 text-to-image generation, creating consistent VGA-style visuals as players embody Malinowski during his Trobriand Islands fieldwork (1915-1918). To address ethical concerns, indigenous peoples appear as silhouettes while Malinowski is detailed, prompting reflection on anthropological representation. Two validation studies confirmed effectiveness: Study 1 with 10 non-specialists showed strong learning outcomes (average quiz score 7.5/10) and excellent usability (SUS: 83/100). Study 2 with 4 expert anthropologists confirmed pedagogical value, with one senior researcher discovering "new aspects" of Malinowski's work through gameplay. The findings demonstrate that AI-driven educational games can effectively convey complex anthropological concepts while sparking disciplinary curiosity. This study advances AI-native educational game design and provides a replicable model for transforming academic texts into engaging interactive experiences.

\end{abstract}



\begin{teaserfigure}
  \includegraphics[width=\textwidth]  {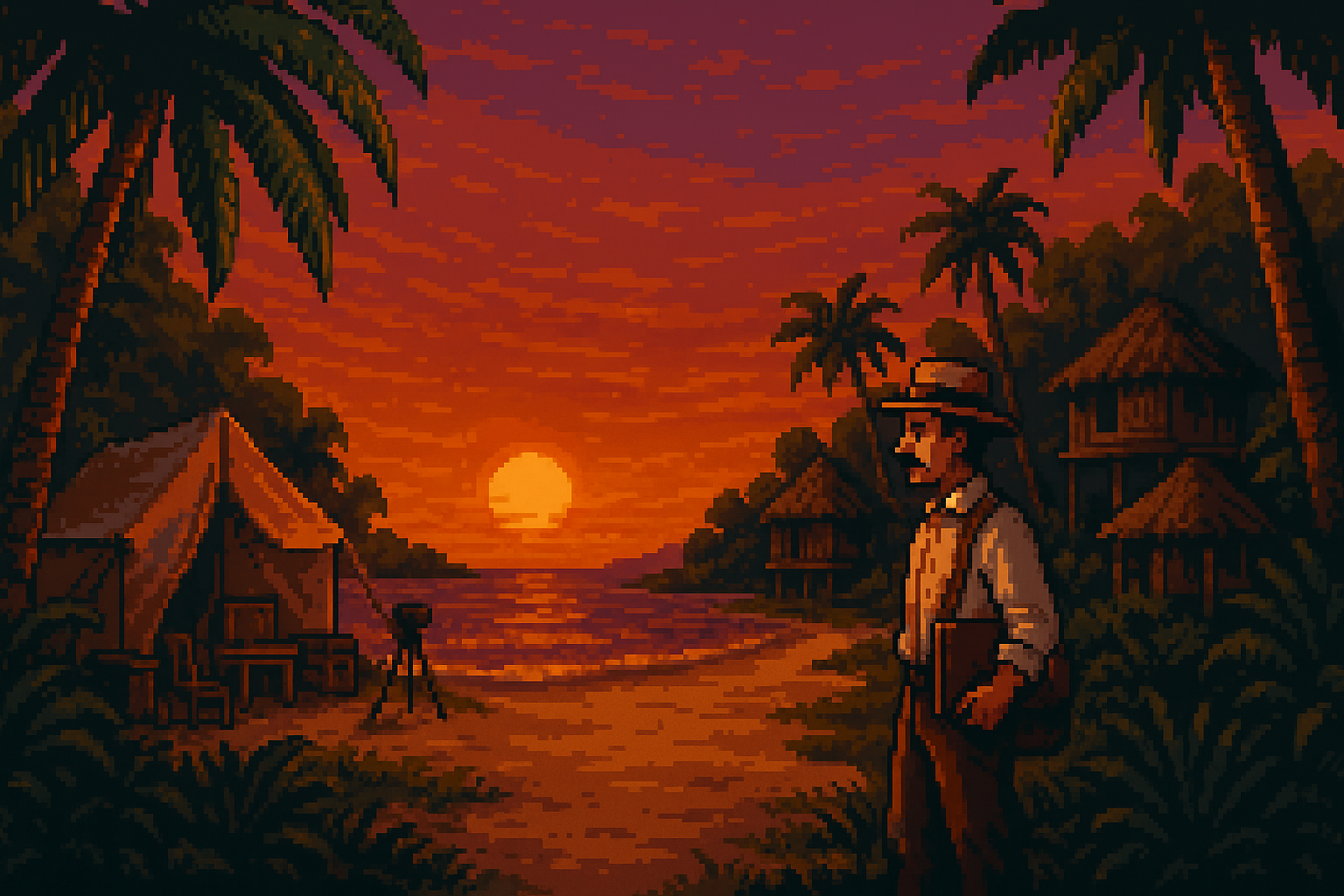}  
  \caption{AI-generated scene from the game 'Malinowski's Lens'}
  \Description{AI-generated scene from the game 'Malinowski's Lens'}
  \label{fig:teaser}
\end{teaserfigure}

\received{20 February 2007}
\received[revised]{12 March 2009}
\received[accepted]{5 June 2009}

\maketitle

\section{Introduction}
The emergence of generative AI has fundamentally transformed human-computer interaction, enabling new paradigms for dynamic content creation and adaptive user experiences. Recent advances in Large Language Models (LLMs) and text-to-image generation have opened unprecedented possibilities for creating AI-native interactive systems where artificial intelligence functions not as a static component but as a responsive partner in real-time content generation and narrative adaptation \cite{sun2023language}. This shift presents unique opportunities to reimagine educational interfaces, moving beyond traditional multimedia presentations toward systems that can dynamically generate contextual content, adapt to user interactions, and provide personalized learning pathways.

Within educational human computer interaction (HCI), there has been growing interest in leveraging AI for interactive learning experiences \cite{bassner2024iris, kharrufa2024potential}. However, most existing approaches focus on conversational interfaces or adaptive assessment systems, with limited exploration of how generative AI can create immersive, visually consistent learning environments. Simultaneously, the field has seen emerging interest in games that convey complex cultural and anthropological concepts—what \citet{hoffmann2025designing} term 'anthrogames'—including titles such as Never Alone (Kisima Ingitchuna) (2014)\footnote{https://www.neveralonegame.com} and Thunderbird Strike (2017)\footnote{https://www.thunderbirdstrike.com}. While these games demonstrate the potential for interactive cultural learning, they rely on pre-created content and traditional game development pipelines, limiting their adaptability and scalability for educational contexts.

This study presents 'Malinowski's Lens', a novel AI-native educational interface that demonstrates a new interaction paradigm: dynamically generated narrative-visual learning experiences. Our system transforms Bronisław Malinowski's classic ethnographic text "Argonauts of the Western Pacific" (1922) \cite{malinowski2013argonauts} into an interactive learning environment through the seamless integration of two AI subsystems: a Retrieval-Augmented Generation (RAG) engine that creates contextually appropriate narratives drawn from the original text, and a DALL·E 3-powered visual generation system that maintains consistent aesthetic coherence while adapting to user actions.

The core HCI contribution lies in our hybrid AI interaction model where users navigate through dynamically generated content spaces. Players embody Malinowski during his Trobriand Islands fieldwork (1915-1918), but unlike traditional games with predetermined storylines, every interaction triggers real-time content generation. The system maintains narrative coherence through RAG-based retrieval while simultaneously generating contextually appropriate visuals in a consistent VGA adventure game aesthetic. This creates what we term "responsive ethnographic simulation"—an interface where user actions dynamically trigger contextually relevant cultural learning moments.

Our system introduces a novel dual-phase interaction paradigm: a fieldwork exploration phase where users engage with AI-generated cultural scenarios and collect ethnographic insights, followed by an academic defense phase where they must articulate their findings to a virtual colleague. This interaction model addresses a key challenge in educational interfaces: balancing exploratory learning with structured knowledge assessment. We deliberately employ visual abstraction—representing indigenous Trobriand Islanders as silhouettes while depicting Malinowski in detail—as both a response to current limitations in AI-generated cultural representation \cite{ghosh2023person, hakopian2024art} and as a design choice that prompts critical reflection on anthropological representation practices. This demonstrates how interface design decisions can address ethical concerns while maintaining educational effectiveness.

To validate our system's effectiveness, we conducted two complementary studies. Study 1 with 10 non-specialist participants demonstrated strong learning outcomes (quiz scores ranged 6-9/10, average 7.5) and excellent usability (SUS: 83/100), with participants expressing increased interest in reading the original text. Study 2 with 4 expert anthropologists confirmed the system's pedagogical value—notably, one senior anthropologist discovered "new aspects" of Malinowski's work through our interface. Expert participants endorsed the system for classroom integration as an active learning supplement.

This research makes three key contributions to HCI: 

First, we introduce Malinowski’s Lens, a novel AI-native educational game that combines real-time narrative generation with consistent visual world-building. The system demonstrates how RAG-based content retrieval can be integrated with generative visual pipelines to create coherent, interactive learning environments.

Second, we advance the understanding of AI-generated educational applications by evaluating Malinowski’s Lens in terms of usability, learning potential, and pedagogical value, providing both user and expert perspectives on its effectiveness.

Third, we contribute design insights derived from participant feedback, which guided iterative refinements of Malinowski’s Lens and highlight broader considerations for future AI-native educational game design.

This work advances our understanding of how generative AI can create novel forms of human-computer interaction while maintaining educational rigor and ethical responsibility. The system demonstrates practical applications for AI-native educational interfaces and provides a foundation for future research in dynamically generated learning environments. 

The remainder of this study is organized as follows. Section \ref{Relatedwork}  reviews related work, while Section \ref{GameDesignAndImplementation} details our system design and implementation. Section \ref{Userstudy} presents the evaluation of our prototype and Section \ref{design} discusses design insights from participants. Then Section \ref{iterative} lays out our iterative prototyping process. Section \ref{limitations} addresses study limitations, Section \ref{Ethics} covers ethical considerations, and \ref{conclusion}  presents our conclusions. Finally, Section \ref{futurework} outlines directions for future work.

\section{Related Work}
\label{Relatedwork}
This section covers related scholarly research on AI-native games and their mechanics, the use of games as teaching tools in the humanities, and RAG-based AI systems for creating educational content.

\subsection{AI-Native Games and their Game Mechanics}
AI-native games represent a departure from traditional approaches to game design. Whereas conventional games employ AI primarily as a supporting technology—for example, to control non-player characters (NPCs), enable procedural content generation, or adjust difficulty levels dynamically \cite{yang2024gpt} — AI-native games place AI models at the center of gameplay and content creation \cite{gallotta2024large}. As industry observers have noted, AI is no longer a peripheral tool in development pipelines but a central component shaping immersion and player experience \cite{forbes2024gaming}. The distinction lies in function: while established studios such as Ubisoft leverage AI to augment gameplay \cite{forbes2024gaming}, for instance, through real-time NPC dialogue generation, AI-native games rely on AI as the primary driver of interaction, narrative, and player engagement.

Several pioneering projects illustrate this shift by positioning AI as a core mechanic rather than an auxiliary feature \cite{sun2023language, sun2025drama}. AI Dungeon\footnote{https://aidungeon.com}, one of the earliest examples, is a text adventure game that uses artificial intelligence to generate dynamic storylines in response to player input. The game demonstrated how AI could serve as the primary content generation engine, creating infinite narrative possibilities through player interaction.

Sun's "1001 Nights" \cite{sun2023language} represents another approach to AI-native design, implementing AI-driven persuasion mechanics where players assume the role of a princess who must convince characters to speak specific words that then manifest changes in the game world. This design showcases how AI can facilitate complex social interaction mechanics that would be impossible with traditional scripted dialogue systems.

A different approach is found in Gandalf \footnote{https://gandalf.lakera.ai/}, which integrates AI security principles into its core gameplay. Across multiple levels, players attempt to coax the AI into revealing hidden passwords, with the system becoming progressively more resistant to manipulation \cite{pfister2025gandalf}. Here, concepts such as adversarial prompting and jailbreak resistance are transformed into mechanics of play, highlighting the potential of AI-native games to make technical issues playable.

Beyond individual titles, research prototypes have also explored AI-native mechanics. The Generative Agents project \cite{park2023generative} populated a sandbox environment with 25 large language model (LLM)-driven characters capable of autonomous communication and evolving social behaviors, creating the impression of a living virtual community. Similarly, CareerSim employed LLMs’ generative and reasoning capabilities to support role-playing simulations of career pathways, extending AI-native mechanics into experiential learning \cite{du2024careersim}.

These developments are not limited to entertainment \cite{lc2023speculative, li2024generative, zhang2024can}. Games powered by generative AI have also been shown to carry educational value. For instance, EcoEcho employed multimodal AI agents to foster sustainability awareness through natural language dialogue and simulated environmental consequences. In a study with 23 participants, the game increased intentions toward sustainable behavior, although shifts in attitudes were less pronounced \cite{zhang2024can}. Likewise, other projects have demonstrated how generative design tools can prototype climate futures \cite{lc2023speculative} and support novice designers in addressing complex sustainability challenges \cite{li2024generative}. Building on this body of work, our study investigates how AI-native games can also foster measurable learning outcomes in an educational context, specifically within social anthropology, highlighting their broader potential as tools for engagement and knowledge transfer.

\subsection{Games as Pedagogical Tools in the Humanities
}

Games increasingly function as legitimate scholarly media across humanities disciplines, offering interactive alternatives to traditional academic discourse. \citet{spring2015gaming} demonstrated how video games could serve as viable scholarly monographs, leveraging the medium's flexibility and multilayered nature to present historical research through "scholarly games". This approach transforms passive academic consumption into active interpretation, as \citet{kapell2013playing} illustrated through their concept of "playable past"—historical video games that engage users as active interpreters rather than passive recipients of information.

Extending beyond historical applications, \citet{reinhard2018archaeogaming}  demonstrated how archaeological knowledge could be effectively communicated through game mechanics, enabling players to engage with methodological processes rather than merely absorbing factual content. \citet{stockhammer2020bronzeon} developed the game 'Bronzeon', a 
civilisation build-up/strategy game
whose rules are all based on the most recent scientific results of a study of the Lech River Valley in Germany. The game informs the player about the hard life in the Bronce age, the importance of
(female) mobility, and the development from copper to complex bronze technology.
 These two examples establish games not simply as entertainment products but as sophisticated tools for academic knowledge production and dissemination, capable of conveying complex scholarly arguments through interactive experience.

Importantly, anthropologists have long recognized the potential of games for conveying disciplinary insights \cite{petridis2021anthropological}, yet—with a few exceptions (e.g., \cite{pia2019digital})—the field has been slow to adopt digital games as scholarly media. Recent work by \citet{hoffmann2024malinowski, hoffmanntext} has begun to bridge this gap by applying LLMs to anthropological material, showing that meaningful games can be generated with minimal hallucinations when combined with RAG \cite{hoffmanntext}. Building on this foundation, our work advances the state of the art in two ways: first, by introducing a more complex game design that integrates narrative, ethics, and interactivity; and second, by pioneering a text-to-image pipeline to create a coherent, playable artefact rather than a simple chatbot. Together, these contributions demonstrate how AI-native game design can transform a canonical academic text into an interactive visually supported experience.

\subsection{RAG in Educational Content Generation}
Artificial intelligence has gained widespread adoption in educational settings (\cite{chen2020application}, \cite{chen2020multi}), with large language models increasingly being explored to enhance learning and teaching processes. However, real-world educational deployment faces significant challenges, particularly the hallucination problem where models generate factually incorrect or misleading information due to their probabilistic nature \cite{ji2023survey}. Additional limitations include the static knowledge embedded within LLMs, which prevents dynamic updates \cite{zhang2024comprehensive}, and their inherent lack of explainability \cite{zhao2024explainability}. To address these shortcomings, researchers have increasingly pursued more reliable solutions, with many turning to Retrieval-Augmented Generation approaches \cite{li2025retrieval}. RAG enhances LLMs by retrieving relevant information from external knowledge bases and incorporating it into the generation process, thereby improving factual accuracy and enabling dynamic knowledge updates \cite{lewis2020retrieval}.

The proliferation of RAG-based solutions in educational contexts has been substantial, as demonstrated by recent comprehensive reviews \cite{li2025retrieval, swacha2025retrieval}. \citet{li2025retrieval} identified four primary applications of RAG for enhancing interactive and personalized learning in education: Educational Q\&A Systems, Educational Chatbots, AI-driven Tutoring Systems, and Adaptive Learning Paths. This systematic categorization illustrates both the widespread adoption of RAG solutions and the core areas where these technologies are making significant educational impact.

Recent implementations exemplify these trends across multiple categories. Notable applications include OwlMentor, an AI mentoring system that supports university students' engagement with scientific literature while demonstrating improved perceived usefulness and alignment with student learning strategies \cite{thus2024exploring}. Automatic lesson plan generation systems have similarly emerged, leveraging RAG architectures to produce curriculum-aligned content \cite{zheng2024automatic}. Furthermore, development frameworks such as Flowwise and LangGraph now provide accessible pipelines for educators to construct domain-specific RAG-based chatbots and learning assistants \cite{csedu25}.

While existing educational RAG applications predominantly focus on traditional chatbot interfaces \cite{swacha2025retrieval}, our work presents a novel approach that transforms classical anthropological literature into immersive gameplay experiences. 'Malinowski's Lens' diverges from conventional question-and-answer paradigms by utilizing RAG technology to create dynamic exploratory environments where players directly experience ethnographic observations. This innovative design maintains scholarly accuracy through faithful source material representation while enabling contextually appropriate content generation that responds to player interactions.

\section{System Design and Implementation}
\label{GameDesignAndImplementation}
This section outlines the design and development of 'Malinowski's Lens'. Section 3.1 details our gameplay design principles and rationale. Section 3.2 describes the game's user interface and Section 3.3 explains the technical implementation.

\begin{figure*}[ht]
  \centering
  \includegraphics[width=\textwidth] 
  {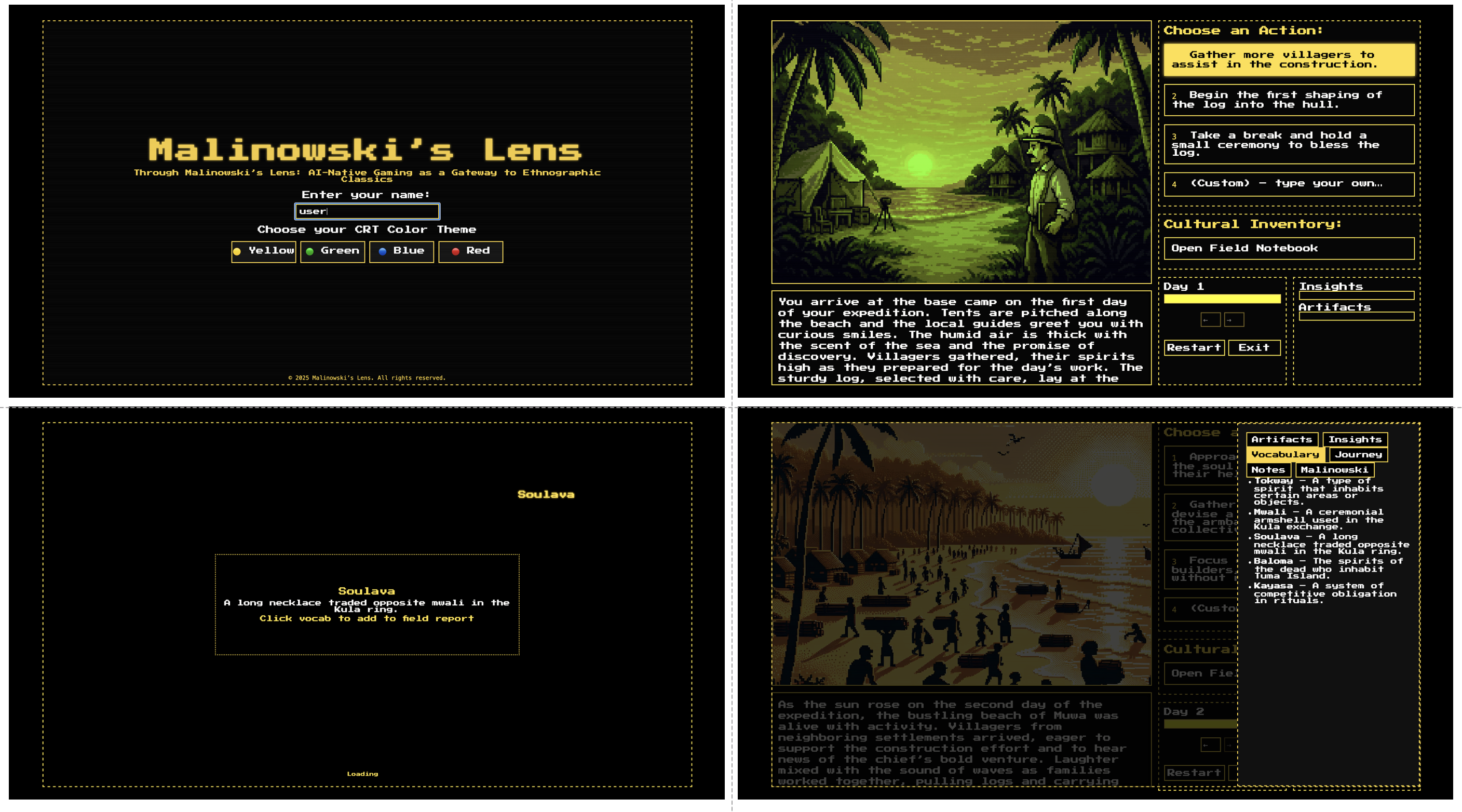} 
  \caption{Sequential screenshots of game components: Start Screen, Interface, Loading Screen, and Itinerary (from top left)}
  
  \label{fig:playthrough}
\end{figure*}

\subsection{Game Design}
\subsubsection{The Player's Journey}
The player begins by establishing their field camp on the Trobriand Islands, where the LLM generates scenarios based on Bronisław Malinowski's seminal work "Argonauts of the Western Pacific" \cite{malinowski2013argonauts}. Each turn in the game presents textual description of the scene with an accompanying image and three AI-generated response options, as well as a custom response option for open-ended exploration.
When the player selects a choice or enters a custom response, a loading screen appears featuring a direct quote from Malinowski's book. During this loading period, local expressions from the original text appear randomly on the screen, which players can collect by clicking on them. When the player clicks on a vocabulary item, its meaning in English is revealed. After a short loading period (approximately 40–50 seconds), the next turn begins, presenting an image, a scene description, and a set of choices (see Figure \ref{fig:playthrough} for the core game components\footnote{At the start of the game, players can select from different color schemes (yellow, green, blue, red) to support accessibility for colorblind users.}).

Through this interactive fieldwork, the player investigates the political, economic, and religious dimensions of Trobriand Island society. Throughout their research, players encounter significant cultural insights—artifacts, local expressions, and important practices—which they can collect by clicking on these elements. Selected cultural items appear in the dedicated \textit{Cultural Inventory}, creating a tangible record of their anthropological discoveries.

Upon gathering four distinct cultural artefacts, the fieldwork phase concludes and the player returns to their university base to face an intellectual challenge from an academic colleague. Before entering this academic defense, players have the opportunity to review their entire fieldwork experience by navigating through the day counters, allowing them to revisit each scene and examine the choices they made during each day of gameplay.

The Academic Defense phase then begins, presenting players with a comprehensive ten-question multiple-choice quiz designed to evaluate their learning throughout the fieldwork experience. This assessment tests their retention of key concepts, cultural understanding, and engagement with the gameplay elements they encountered. At the conclusion of this intellectual exchange, a scoreboard displays the player's performance, providing immediate feedback on how well they absorbed and retained the lessons from their anthropological journey\footnote{A video demonstration of 4 minutes of gameplay as well as the quiz is available at \href{https://www.dropbox.com/scl/fo/jlzar4mmn46wajh1eqrtr/AB7wsyBHwPegjvKwLYw7H7Q?rlkey=41ffkhfsyioq4eo0yun0mc1jz&st=ppck3a3v&dl=0}{\textcolor{blue}{dropbox}}, anonymized to protect participant privacy.}.

\subsubsection{Ethnographic Collection: Discovering Cultural Treasures}
While exploring the Trobriand Islands, players engage with two primary activities. First, they navigate the political, economic, and religious dimensions of island life through scenario-based narratives derived from Malinowski's "Argonauts of the Western Pacific" via the RAG-system. Second, and equally important, they must identify and collect significant cultural elements—tangible objects, conceptual insights, and local expressions.

These ethnographic treasures are automatically identified by a parser that analyzes the AI-generated content. When encountered, these cultural elements are added to the player's inventory. This collection mechanic creates a purposeful progression system while reinforcing key anthropological concepts.

Drawing inspiration from contemporary games like "1001 nights" \cite{fu2025like} and the culturally immersive "Never Alone" (Kisima Inŋitchuŋa), this feature transforms passive reading into active discovery. By establishing the clear objective of gathering four distinct cultural artefacts before advancing to the Academic Defense phase, the mechanic maintains player engagement while structuring the ethnographic learning journey through meaningful goals and tangible progress markers.

\subsubsection{Dynamic Environment Design: Responsive Visuals with Ethical Considerations}    
To deepen immersion and engagement, we designed a dynamic environment system that generates responsive visuals in real time. Narrative scenarios, produced via the RAG system, are distilled into concise prompts for DALL·E 3, which in turn generates VGA-style background images (See Appendix \ref{imagegeneratonprompt} for an example). This design operationalizes principles of emotional game design \cite{isbister2016games} by aligning visual feedback directly with narrative progression, thereby reinforcing player–story connection.

Our choice of a consistent early-1990s VGA adventure game aesthetic was deliberate: it provides a cohesive visual identity, evokes nostalgia associated with exploratory gameplay, and differentiates the educational experience from contemporary photorealistic genres. Importantly, our visual representation strategy addressed the ethical complexities of depicting indigenous communities. To mitigate risks of misrepresentation by Western-trained AI models, we rendered Trobriand Islanders as silhouettes while presenting Malinowski himself in detail. This technique echoes initiatives in art such as Silvia Kolbowski's explorations of postcolonial shadows\footnote{Silvia Kolbowski is noted for her use of shadows and the deliberate absence of human figures. In works such as Monumental Prop/portions, she removes the body from the image, leaving only cropped, expressive shadows (see https://www.printedmatter.org/catalog/8949/)} but also draws from \citet{lapensee2021rivers} research on indigenous representation in digital media, which emphasizes respectful portrayal while acknowledging the inherent complexities of cross-cultural depiction. In Section \ref{generalResults}, we report how players interpreted this visual strategy and how it shaped their sense of immersion and authenticity.

\subsubsection{Academic Defense: From Fieldwork to Scholarly Scrutiny}
The second phase of the game, the Academic Defense, models the process of defending fieldwork in scholarly contexts. Once players have gathered four cultural artefacts, they enter a quiz-based dialogue where an AI agent plays the role of a critical academic peer. The system generates ten multiple-choice questions: one quotes Malinowski’s Argonauts, one addresses the theoretical basis of the game, three test vocabulary encountered during pauses, two focus on collected artifacts, and the remaining three probe narrative comprehension.

This mechanic creates a payoff structure for the fieldwork phase by requiring players to actively demonstrate retention and interpretation. It echoes existing platforms such as Kahoot! \cite{wang2015wear}, but differs in its dynamic tailoring of questions to player-generated gameplay histories. The approach situates assessment not as an external add-on but as an embedded continuation of the anthropological journey. We analyze in Section \ref{generalResults} how players responded to this academic defense and whether it supported deeper engagement with the fieldwork phase.

\subsection{User Interface}
\begin{figure}[H]
  \centering
  \includegraphics[width=\columnwidth]
  {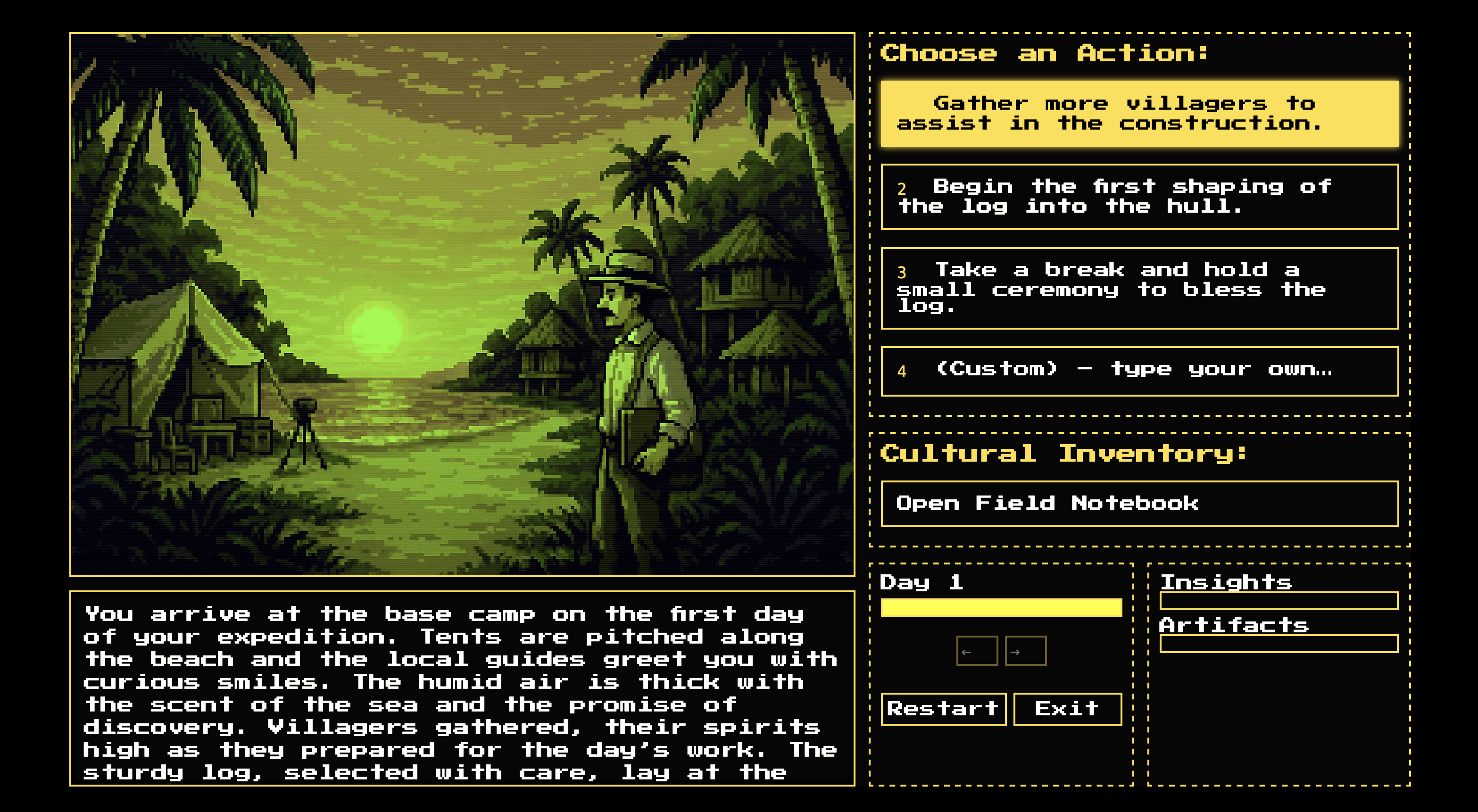}
  \caption{UI Interface}
  \label{interface}
\end{figure}

The user interface (UI) of Malinowski’s Lens is designed to balance narrative immersion with interactive learning, drawing inspiration from early text-adventure games while integrating affordances of modern educational technology. The screen is divided into two primary zones (Fig. \ref{interface}): a left-side narrative space and a right-side interaction panel. The left panel features a pixel-based scene visualization, which dynamically changes based on the player's choices. Below this image, a contextual text box presents a brief scene description grounded in retrieved ethnographic content. This visual-verbal pairing aims to support multimodal learning and narrative immersion, reinforcing the connection between player decisions and cultural context.

The right panel serves as the primary interaction zone. It presents four response options: three AI-generated choices and one open-ended input field that allows players to type their own response. This hybrid model supports both structured learning and free-form exploration, inviting players to co-construct meaning through natural language interaction. Beneath the interaction panel, additional UI elements track player progress. These include a cultural inventory (listing discovered insights and artifacts), a day counter, and real-time metrics indicating the number of collected expressions and cultural items. These components function both as feedback mechanisms and as learning reinforcements, visually marking progress through the gameplay.




\subsection{Technical Architecture of the Interactive Storytelling System}
\begin{figure*}[!htbp]
  \centering
  \includegraphics[height=0.46\textheight]{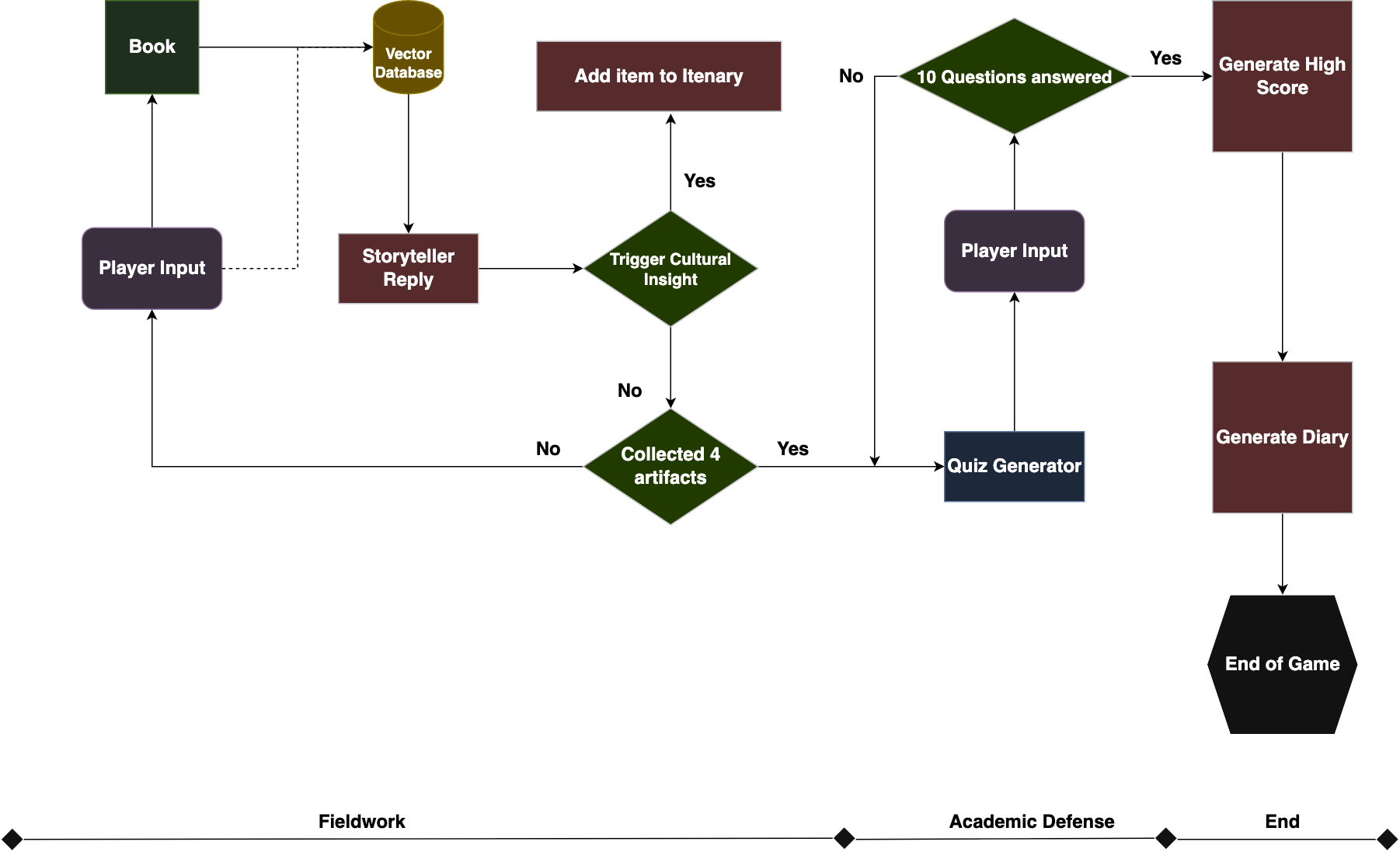}
  \caption{Technical Workflow}
  \label{fig:technicalWorkflow}
\end{figure*}

The technical foundation of Malinowski’s Lens was developed to meet three core requirements: (1) ensuring fidelity to ethnographic source material, (2) enabling dynamic and adaptive gameplay, and (3) maintaining extensibility for future learning contexts. To address these goals, we implemented a modular Retrieval-Augmented Generation pipeline integrated with a layered application architecture (Fig. \ref{fig:technicalWorkflow}).
At the knowledge retrieval layer, the system ingests Malinowski’s Argonauts of the Western Pacific (1922) \cite{malinowski2013argonauts}. The text is pre-processed into high-dimensional embeddings using the OpenAI text-embedding-3-small model\footnote{https://platform.openai.com/docs/models/text-embedding-3-small}, and stored in a ChromaDB vector database for efficient retrieval. When players interact with the game, the retrieval engine queries ChromaDB\footnote{https://www.trychroma.com} to surface semantically relevant excerpts. This architecture constrains narrative generation to authentic ethnographic passages, reducing hallucinations and ensuring cultural and textual fidelity.

The narrative generation layer integrates retrieval with generative modeling. Retrieved passages are dynamically composed with structured prompts orchestrated through LangChain\footnote{https://www.langchain.com}, and then passed to an LLM (GPT-4o). This setup ensures that generated content remains contextually grounded while preserving the flexibility required for interactive storytelling. Prompt templates explicitly encode pedagogical goals, guiding the model to produce both culturally accurate and educationally coherent scenarios (See Appendix \ref{imagegeneratonprompt}).

The application layer is built on a FastAPI\footnote{https://fastapi.tiangolo.com} backend, which manages retrieval orchestration, prompt assembly, and model interaction. The backend exposes endpoints to a React-based single-page frontend built with Vite\footnote{https://vite.dev}. The frontend delivers the pixel-based text-adventure interface, handling real-time rendering of images, display of retrieved ethnographic text, and player interaction (including multiple-choice selections and free-text input). Figure \ref{fig:technicalWorkflow} illustrates the complete technical workflow, demonstrating the integration between vector-based retrieval, contextual generation, and interactive presentation layers that collectively enable the transformation of classical ethnographic texts into dynamic, educationally-focused gaming experiences.

\section{User Study}
\label{Userstudy}
We conducted a formative evaluation to assess how effectively 'Malinowski's Lens' conveys key anthropological concepts from 'Argonauts of the Western Pacific' (e.g., the Kula ring\footnote{The Kula ring is a ceremonial exchange system where Pacific islanders trade shell ornaments in circular patterns to create social bonds beyond economic value.}, participant observation methodology\footnote{Participant observation is an ethnographic method where researchers live within a community to study it from the inside by both participating and observing.}), its potential to generate interest in anthropological scholarship among non-specialists, and to identify areas for prototype refinement. Given the novel application of AI-assisted narrative generation in educational game design for anthropological content, this initial evaluation adopts an exploratory approach to understand both user experience and disciplinary validity. Therfore
our study employs a mixed-methods design incorporating think-aloud protocols, performance assessments, and semi-structured interviews across two distinct participant groups. This approach enables us to examine both the game's pedagogical effectiveness for its intended audience and its scholarly accuracy from expert perspectives, providing foundational insights for future iterations and larger-scale validation studies.

\subsection{Study Design Rationale}

Our study design incorporates two distinct participant groups to comprehensively evaluate 'Malinowski's Lens' from complementary perspectives. We recruited 10 general users with minimal anthropological background alongside 4 professional anthropologists, creating a mixed-methods approach that addresses both pedagogical effectiveness and disciplinary authenticity.

\textbf{General User Group (n=10):} The primary focus on general users aligns with the game's core objective of making anthropological concepts accessible to broader audiences. These participants, recruited through snowball sampling and social media networks, represent the game's intended user base—individuals with little to no prior knowledge of social anthropology who have never engaged with 'Argonauts of the Western Pacific'. This group allows us to evaluate whether the game successfully translates complex anthropological theory into an engaging, comprehensible interactive experience. Their responses provide crucial insights into the game's pedagogical effectiveness, user experience design, and potential to generate interest in anthropological scholarship among non-specialists.

\begin{table*}[t]
    \caption{Participant Demographics, Background, and Experience}
    \label{tab:participant_info_compact}
    \begin{tabular}{c c c c c c}
    \toprule
    \textbf{ID} & \textbf{Age} & \textbf{Gender} & \textbf{Education / Occupation} & 
    \textbf{Anthropology Knowledge} & \textbf{Game Experience} \\
    \midrule
    P1  & 30  & M & Master’s / AI Engineer             & None   & Once a month \\
    P2  & 44  & M & Master’s / AI Engineer             & Little & Once a year \\
    P3  & 52  & M & PhD / Anthropologist               & Expert & Rarely \\
    P4  & 29  & M & PhD Student / AI Engineer          & None   & None \\
    P5  & 40  & F & PhD / Anthropologist               & Expert & None \\
    P6  & 42  & M & PhD / Anthropologist               & Expert & None \\
    P7  & 35  & F & PhD / Engineer                     & Little & None \\
    P8  & 29  & M & Master’s / Computational Scientist & Little & Once a month \\
    P9  & 38  & M & PhD / Anthropologist               & Expert & Rarely \\
    P10 & 46  & F & Master’s / Light Designer          & None   & None \\
    P11 & 41  & F & PhD / Teacher                      & None   & Rarely \\
   
    P12 & 34 & F & PhD / Computer Scientist           & None   & Rarely \\
    P13 & 48  & F & Master’s / Journalist              & None   & None \\
    P14 & 51  & M & Master’s / Marketing               & None   & Rarely \\
    \bottomrule
    \end{tabular}
\end{table*}

\textbf{Expert User Group (n=4):} The inclusion of professional anthropologists serves as a critical validation mechanism for the game's representation of Malinowski's work and broader anthropological concepts. These participants can assess whether key theoretical frameworks—particularly participant observation methodology and the Kula ring system—are accurately conveyed without oversimplification or misrepresentation. Their expertise enables evaluation of the game's scholarly integrity and identification of potential conceptual errors that general users might not detect. Additionally, anthropologist participants can provide informed perspectives on how the game might be integrated into educational contexts and suggest refinements that enhance both accuracy and pedagogical value.

\textbf{Methodological Considerations: }This dual-group design recognizes that educational games targeting academic content must satisfy two distinct but interconnected criteria: accessibility for novice learners and credibility among domain experts. By incorporating both perspectives within a single study, we can identify tensions between simplification for accessibility and maintenance of conceptual accuracy, ultimately informing design decisions that optimize both dimensions. The smaller expert group size reflects practical constraints in recruiting professional anthropologists while still providing sufficient data for meaningful analysis of disciplinary perspectives.

\subsection{Study 1: General Users}
\subsubsection{Recruitment}
We recruited 10 participants representing the game’s target audience through snowball sampling and social media networks (see Table \ref{tab:participant_info_compact}). These participants came from diverse professional backgrounds, including AI engineering (P1, P2, P4), computational sciences (P7, P8), computer sciences (P12), light design (P11), teaching (P10), journalism (P13), and marketing (P14). All held advanced degrees (six Master’s and four PhD degree holders), but none had formal training in anthropology. Consistent with our recruitment goals, all reported minimal or no prior knowledge of social anthropology and had never engaged with Argonauts of the Western Pacific, ensuring an authentic evaluation of the game’s introductory pedagogical function. Participants volunteered without monetary compensation and provided informed consent following a briefing on study objectives and procedures.
\begin{table*}[t]
    \caption{SUS Test Questions: Statistical Analysis by Item (7-Point Likert Scale)}
    \label{SUSquestions}
    \begin{tabular}{ccp{1cm}ccc}
    \toprule
    \textbf{Question} & \textbf{Statement} & \textbf{Mean} & \textbf{Median} & \textbf{Std dev} \\
    \midrule
    1 & I think that I would like to use this game & 5.79 & 6.0 & 1.58 \\
    2 & I found the game unnecessarily complex. & 1.86 & 2.0 & 0.77 \\
    3 & I thought the game was easy to use. & 6.36 & 6.5 & 0.74\\
    4 & I think that I would need the support of a \\
     & technical person to be able to use this game.  & 1.00 & 1.00 & 0.00 \\
    5 & I found the various functions in this game were well integrated. & 4.57 & 4.00 & 1.65 \\
    6 & I thought there was too much inconsistency in this game.  & 3.21 & 3.0 & 1.48 \\
    7 & I would imagine that most people would learn \\
     & to use this game very quickly.& 6.71 & 7.0 & 0.47 \\
    8 & I found the game very cumbersome to use. & 1.43 & 1.0 & 0.51 \\
    9 & I felt very confident using the game. & 5.86 & 6.0 & 1.1 \\
    10 & I needed to learn a lot of things before I could get going with this game. & 1.79 & 2.0 & 1.05 \\
    \bottomrule
    \end{tabular}
\end{table*}

\subsubsection{Procedures}: 
The study protocol consisted of four sequential phases designed to capture both behavioral and attitudinal responses to the game. All sessions were conducted in a controlled laboratory environment where participants accessed 'Malinowski's Lens' on standardized laptop configurations while employing think-aloud protocols to externalize their cognitive processes throughout the gameplay experience.

\textit{Phase 1 - Introduction (5 minutes):} Following informed consent procedures, participants received standardized instruction on game mechanics, interface navigation, and the think-aloud protocol expectations. This orientation ensured consistent baseline understanding across participants.

\textit{Phase 2 - Exploratory Gameplay (30 minutes):} Participants engaged with the game's primary exploration mechanics, investigating the virtual Trobriand Islands through AI-generated narrative paths or custom exploration choices. The 30-minute duration was selected to allow sufficient engagement with core game mechanics while maintaining participant attention. Participants who accumulated adequate cultural artefacts to naturally progress to the academic defense phase continued seamlessly, while those who had not reached this threshold after 30 minutes were manually advanced to maintain consistent study timing. The large majority of participants completed the exploration phase within the allotted time, with only a small minority requiring manual advancement—primarily due to extended discussions of game mechanics, design considerations, or emergent ideas that arose in conversation with the authors during gameplay.

\textit{Phase 3 - Knowledge Assessment (10 minutes):} All participants completed a 10-item multiple-choice assessment evaluating comprehension of key anthropological concepts encountered during gameplay. The assessment consisted of LLM-generated questions designed to test understanding of the specific anthropological content presented during the game session. Each question presented four response options with a single correct answer, scored using a binary point system (1 point per correct response, 0 for incorrect).

\textit{Phase 4 - Post-Study Evaluation (15 minutes):} Participants completed the standardized System Usability Scale (SUS) instrument to assess perceived usability (See Table 3), followed by a semi-structured interview examining: (1) perceived learning outcomes related to Malinowski's work, (2) changes in interest toward anthropological scholarship, (3) overall gameplay experience, and (4) suggestions for prototype improvements (See Appendix Table \ref{survey}). Interviews were led by the key author while a second author of this study took notes with participant consent for subsequent thematic analysis\footnote{We documented each playtest session with images and detailed notes, which are stored in the study's repository at https://github.com/XXX in anonymized form and are available upon request.}.

\subsubsection{Results}
\label{generalResults}

All ten participants from the general user group completed the gameplay and assessment tasks of the game. The results suggest that, at the most general level, 'Malinowski's Lens' appears to help convey complex anthropological concepts to a lay audience, as well as raise interest in both Malinowski's book as well as social anthropology as a whole. The participants of the study suggested different future improvements for the game prototype.

\textit{Quantitative Findings:}
We evaluated the usability of the game using the standardized System Usability Scale (SUS) \cite{brooke1996sus}, administered to all 10 general user participants. Each participant responded to the ten standard items, originally rated on a 7-point Likert scale and subsequently normalized to the conventional 5-point format\footnote{We used a 7-point Likert scale to allow for more nuanced responses, then normalized results back to the standard 5-point SUS format for analysis.}. Table \ref{SUSquestions} shows the questions along the mean, median and standard deviation for each question. Table \ref{resultandindividualSUS} shows how individual participants rated each question.

The resulting overall SUS score was 83 out of 100, which places the system in the “excellent” usability range according to the SUS-Bangor adjective rating scale \cite{bangor2009determining}. This suggests that players generally found the game accessible and easy to use, meeting expectations for an experimental educational game prototype.

To assess learning outcomes, we administered a 10-item knowledge quiz immediately following gameplay. Scores ranged from 6 to 9 correct answers (out of 10), with a mean of 7.5, a median of 7.5 and a standard deviation of 0.81 (See Appendix \ref{individualquizzscores}). This performance indicates that participants successfully acquired and retained key cultural concepts presented during play. While variation in individual scores suggests differing levels of recall, all participants demonstrated comprehension above chance.

Together, the SUS results and quiz outcomes provide quantitative evidence that the system is both usable and effective as a learning tool.


\textbf{Qualitative Feedback On Players’ Perception After The Gameplay}
The post-gameplay interviews revealed distinct patterns in how participants engaged with and perceived Trobriand cultural elements through the game interface. Overall, participants P1, P2, P3, P4, P7, P8, and P10–P14 reported that they found the game enjoyable and fun. However, P4, a native French speaker who also spoke fluent English, noted encountering a repetitive sequence that lasted two days, describing it as 'piétiner sur place' (taking small steps on the spot without moving forward).

Participants P1, P2, and P4 not only engaged with the narrative and choice-based interactions but also experimented with the open text input as a way to “shortcut” progression. Each attempted to request cultural artefacts directly—P1 asked for the mwali\footnote{Mwali are shell armbands or arm ornaments made from large white shells}, P2 suggested to “rob the yam and run away,” and P4 requested a coconut—items that were otherwise meant to be earned through gameplay. Notably, P4 proposed turning this behavior into a mechanic itself, suggesting that an agent could detect such shortcut attempts and respond playfully, thereby transforming the act of “gaming the system” into an intentional and engaging feature.  

P1 demonstrated particular fascination with the economic and social structures of Trobriand society, specifically mentioning "the barter system, rituals, hierarchies" as aspects that sparked curiosity. The participant was intrigued by the representation of indigenous peoples as silhouettes and expressed motivation to understand "world without technology, how do they go on with their daily lives compared to modern life". P2 focused more on community dynamics, highlighting "the ritual, relationship with their community and environment and living" as compelling elements. Both participants found the descriptive text style engaging, with P2 noting that "even if it is tense, style is descriptive, eloquent in the game", which contributed to their interest in Pacific cultures more broadly.

Participants shared mixed feedback on the game’s visual and interactive design. P1 highlighted certain visual details—such as “the coconut tree” and the use of shells as ornaments in fieldnote books—as effective in enhancing cultural immersion. P2 appreciated the generated images, while P7 noted a mismatch between the quality of text and images, suggesting that the visuals could provide greater educational value.

At the same time, participants pointed out areas for improvement in the interface. P2 recommended creating a cleaner layout by avoiding cramped groupings of elements like “artifacts” and “vocabulary” and remarked that the open field notebook appeared cluttered and unappealing. P1 proposed adding audio narration to enrich the experience, while also expressing satisfaction with the 2D visual style—particularly how illustrations such as the sunrise aligned well with the accompanying text.

P1 and P8 added that the visual depiction of the game always included diversity, and complained that at times their visual descriptions diverged from the trobriand culture \footnote{While this requires additional investigation, the phenomenon likely stems from DALL-E 3's apparent implementation of a diversity filter that activates in response to specific prompt terminology, as documented by \citet{baum2024rendering}}. Some technical issues were also observed, including occasional text repetition and the need for better guidance systems to keep players within appropriate cultural contexts during gameplay. 


\textit{Emerging Curiosity about Anthropological Works:}

The playtest results revealed that Malinowski's Lens successfully generated interest in both the source material and broader anthropological inquiry. Participants P1, P2, P4, P11, P12, P13 and P14 expressed a desire in the post-play interviews to read Argonauts of the Western Pacific following their gameplay experience, suggesting the game effectively functioned as a gateway to Malinowski's ethnographic work. 
P12 expressed curiosity about the cultural artefacts that have to be collected in the game. "These artefacts made me more curious about the book, and I d love to read it". P4 even said that "he plans to buy the book, as the game really got me curious. I thought that relationships between people were much more violent, but this game portrayed them living harmoniously and living in a peaceful setting. I definitely want to buy the book and find out more about life there".  This finding indicates that interactive engagement with anthropological concepts through gameplay can create pathways to academic literature that might otherwise seem inaccessible to general audiences. 

Beyond interest in the specific source material, the gameplay experience appeared to catalyze broader curiosity about anthropological studies and regional cultures. P2 articulated a specific desire to learn more about Pacific cultures, stating that the game "may spark interest to take a look on those cultures or people living in pacific ocean", while P1 developed a more general interest in anthropological approaches to understanding "non-technical cultures", expressing curiosity about "world without technology, how do they go on with their daily lifes compared to modern life!". Both participants acknowledged that their single gameplay session felt introductory rather than comprehensive, with P2 noting that he "needed more play to fully understand the culture" and describing the experience as "more like an introduction". P11 added that the "game could do more, it could deliver more insights" specifically referring to the need to incorporate the colonial context as discussed in the design suggestions. This suggests that while the game successfully ignited curiosity, participants recognized the complexity of cultural understanding and desired deeper engagement with both the interactive medium and the anthropological concepts it presented.

\subsection{Study 2: Senior Anthropologists}
\subsubsection{Preparation:} We recruited four senior anthropologists who regularly teach undergraduate and graduate courses at European universities. Each anthropologist followed the same evaluation procedure described previously. Additionally, we conducted open discussions with participants to gather their perspectives on several key aspects: the game's entertainment value, the quality of both visual and narrative outputs, concerns about potential student dependency on the tool, and their willingness to incorporate it into their classroom instruction.

\subsubsection{Findings:} 
The authors explored participants' experiences with the game, examining both its entertainment value and educational potential. Participant responses to the game's enjoyability were mixed. P3 found it "definitely fun", while P5 provided a more critical assessment, describing it as "intriguing but not fun—for that it would have to provide more agency in the game". P3, a senior anthropologist, noted that the game prompted him to reread Malinowski's original text, through which "he discovered some new aspects of the book that he wasn't aware of before", demonstrating the game's potential to enhance understanding even among subject matter experts.

Participants viewed the game primarily as an educational supplement rather than a standalone learning tool. P5, a senior anthropologist, characterized it as "an introduction to the book", acknowledging that "I don't think anyone will become an anthropologist from playing such games, as this is virtual fieldwork, it helps to get people more interested in the subject". When addressing concerns about student dependency on digital tools, P5 emphasized the game's active learning requirements: "I'm not worried that students get dependent on the game. It requires some effort on behalf of the student. It's not just passive consumption. It makes you think. Because of that I don't think it will replace reading practices."

All four participants stressed the importance of proper pedagogical integration. P5 expressed enthusiasm for classroom implementation, stating she would "definitely integrate the game into her teachings in class, but not without embedding it into a pedagogical context". She viewed the game as "a great starting tool for people to get into the book". Similarly, P6 indicated he would "definitely use it for his classroom teachings albeit only as a bonus to the reading of the book", expressing concern that without proper framing, gaming might substitute for reading the original text.

\begin{table*}[t]
    \caption{Individual Question Responses (1-7 Scale) and Calculated SUS Scores by Participant}
    \label{resultandindividualSUS}
    \begin{tabular}{ccp{1cm}cccccccccccc}
    \toprule
    \textbf{Question} & \textbf{P1} & \textbf{P2} & \textbf{P3} & \textbf{P4} & \textbf{P5} & \textbf{P6} & \textbf{P7} & \textbf{P8} & \textbf{P9} & \textbf{P10} & \textbf{P11} & \textbf{P12} & \textbf{P13} & \textbf{P14}\\
    \midrule
    1 & 6 & 6 & 7 & 7 & 3 & 2 & 7 & 7 & 5 & 6 & 5 & 7 & 6 & 7 \\
    2 & 2 & 2 & 1 & 1 & 4 & 2 & 2 & 2 & 2 & 1 & 2 & 1 & 2 & 2 \\
    3 & 7 & 5 & 5 & 6 & 6 & 7 & 7 & 7 & 6 & 7 & 6 & 7 & 7 & 6\\
    4 & 1 & 1 & 1 & 1 & 1 & 1 & 1 & 1 & 1 & 1 & 1 & 1 & 1 & 1 \\
    5 & 7 & 4 & 3 & 7 & 3 & 2 & 4 & 4 & 3 & 7 & 4 & 5 & 5 & 6 \\
    6 & 3 & 5 & 2 & 5 & 4 & 4 & 1 & 6 & 4 & 2 & 3 & 2 & 2 & 2 \\
    7 & 7 & 6 & 7 & 7 & 6 & 7 & 6 & 7 & 6 & 7 & 7 & 7 & 7 & 7  \\
    8 & 1 & 2 & 1 & 1 & 2 & 1 & 2 & 1 & 2 & 1 & 1 & 1 & 2 & 2 \\
    9 & 7 & 6 & 7 & 7 & 4 & 5 & 6 & 7 & 4 & 6 & 5 & 7 & 5 & 6 \\
    10 & 1 & 2 & 1 & 5 & 1 & 2 & 2 & 1 & 2 & 2 & 2 & 2 & 1 & 1 \\
    \midrule
    SUS & 93 & 75 & 88 & 85 & 67 & 72 & 87 & 85 & 72 & 93 & 80 & 93 & 87 & 90 
    \end{tabular}
\end{table*}

\section{Design Insights and Directions}
\label{design}
Through our user studies and post-interaction interviews, we identified key design insights that advance our understanding of AI-native educational interfaces. These findings reveal critical considerations for designing effective human-AI collaborative learning systems and provide transferable principles for similar interactive applications. This section presents five distinct design directions that emerged from participant feedback, each contributing to broader HCI knowledge about AI-mediated educational experiences.

\subsection{Contextual Transparency in AI-Generated Narratives}
Expert participant P3, a senior anthropologist, identified a fundamental challenge in AI-native educational systems: the need for users to understand the historical and methodological context underlying AI-generated content. P3 suggested incorporating pre-fieldwork historical context about Malinowski's circumstances before arriving on the islands, noting that "certain expressions that the LLM rightly reads from the original text, have to be seen in its colonial context". This insight highlights a critical design principle for AI-educational interfaces: contextual transparency—users need explicit understanding of both the source material and the AI system's relationship to that material to properly interpret generated content. This finding extends beyond our specific application to any AI system that transforms historical or culturally sensitive texts. The design implication suggests that AI-native educational interfaces should include contextual framing mechanisms (such as introductory content or companion materials) that help users critically evaluate AI-generated narratives within their appropriate historical and methodological frameworks.


\subsection{Interface and Immersion Enhancements}
Participants identified several opportunities to enhance the game's interface design and immersive qualities through audiovisual and interaction improvements. P1 suggested integrating weather simulation into the experience, noting that while "the sun images fit with textual descriptions, it must have rained sometimes". This observation led to recommendations for expanding the prompt system to not only describe different weather conditions across different days but also display them visually and incorporate corresponding audio elements such as storm sounds, rain, and thunder to create a more complete sensory experience. P9, an expert anthropologist, even suggested to include soundscapes to make the game more immersive. 

P2 identified issues with information density and narrative transparency that could improve user experience. Regarding the game's pacing, P2 observed that "the itinerary is too cramped; there's too much in it" and recommended making it more streamlined. Additionally, P2 suggested enhancing narrative transparency by making it clear from the onset that the narrative is generated based on Malinowski's book. This could be achieved by adding an explanatory sentence on the start page where rules are explained, acknowledging that while the LLM attempts to stay close to the source material, there is no 100 percent guarantee of complete fidelity to the original text.

P4 proposed interface enhancements focused on cultural accessibility and visual progression. These included implementing a vocabulary lookup feature for local idioms encountered in the text, improving interface padding for better visual hierarchy, and creating a progressive revelation system for cultural artifacts that would initially display them in black and white before transitioning to full color, potentially creating a sense of discovery and understanding as players engage more deeply with the cultural content.

Participants P4, P12, and P14 proposed enhancements to the quiz experience. P4 and P12 suggested enabling users to request hints from the LLM to assist with answering questions. P14 additionally recommended introducing a “wild card” feature that would eliminate two of the four options, noting that “this would make it even more fun”.

\subsection{Content Moderation in Open-Ended AI Interactions}

Multiple participants (P1, P2, P4, P6, P9, P10, P13) raised concerns about users exploiting the custom input feature to submit inappropriate content. P6 noted "you never know, there's always someone who writes something embarrassing, it might be something racist, hateful, so content moderation is very important". This insight reveals a critical challenge in designing AI systems that balance user agency with safety considerations.
The design principle that emerges is proactive safety through multilingual content moderation—AI-native educational systems, especially when portraying vulnerable or marginal groups, must implement robust content filtering while preserving the open-ended interaction that makes them educationally valuable. This finding has broad implications for any educational AI system that allows user-generated input, suggesting the need for robust content moderation without compromising user experience.


\subsection{Performance Expectations in AI-Native Interfaces}
Participants expressed divergent views on system response times, revealing important insights about user expectations in AI-native systems. P2 found inference speed problematic, noting "people are used to image generation being quickly" while P4 viewed slower generation as "an integral part of the retro-game aesthetic". This disagreement illuminates the contextual nature of performance expectations in AI interfaces—users' tolerance for latency depends significantly on the framing and aesthetic context of the interaction.
This insight suggests that AI interface designers should consider how aesthetic and contextual framing affects user tolerance for AI processing delays, potentially using waiting periods as opportunities for anticipation-building rather than sources of frustration.


\subsection{Accessibility and Deployment Considerations}

\begin{table}[htbp]
  \caption{Development and Playtest Costs (Rounded)}
  \label{costs}
  \begin{tabular}{l r}
    \toprule
    \textbf{Phase / Item} & \textbf{Cost (EUR)} \\
    \midrule
    \multicolumn{2}{l}{\textit{Development Phase}} \\
    \quad Text Generation & 5  \\
    \quad Image Generation & 18 \\
    \midrule
    \multicolumn{2}{l}{\textit{Playtest Phase}} \\
    \quad Text Generation  & 1  \\
    \quad Image Generation & 9 \\
    \midrule
    \textbf{Total} & \textbf{33} \\
    \bottomrule
  \end{tabular}
\end{table}

Participants P2 and P4 emphasized the importance of system accessibility, with P2 expressing desire for offline capability: "Would be great to have it on my laptop and I can play without needing access to the internet". Combined with the cost considerations shown in Table \ref{costs}, this feedback highlights the accessibility-performance trade-off inherent in AI-native educational systems.
While our prototype costs remained manageable (€33 total for development and testing), widespread classroom deployment would require more cost-effective solutions. This insight reveals a fundamental tension in AI-educational interface design: cutting-edge AI capabilities often require cloud-based proprietary models, but educational accessibility demands local, affordable deployment options.

\section{Iterative Development}
\label{iterative}
Following our collection of diverse design feedback, we were able to directly implement four key improvements to the prototype.    

\textbf{Enhanced Scene Understanding: }Participants P3 and P4 highlighted the need for a lookup mechanism to clarify unclear terms within scene narrative descriptions. Rather than creating a static dictionary with predetermined keywords, we developed a more flexible solution that leverages our underlying GPT-4o model through a RAG system to provide contextual information about the book. This implementation is demonstrated in Figures 5 and 6 (See appendix \ref{additionalscreenshots}).

\textbf{Interactive Book Consultation:} We also introduced a new chat feature called "Ask about Book" (shown in Figure 7, See appendix \ref{additionalscreenshots}), which enables users to inquire directly about specific content from the original text. This addition transforms the prototype into a tool that supports immersive literary study, allowing players to deepen their engagement with the source material during gameplay.

\textbf{Contextual Game Introduction:} Three senior anthropologists (P3, P5, and P9) suggested incorporating a more contextual introduction to the game. In response, we developed a short introductory text that is now displayed at the beginning of the game. This introduction explains how Malinowski arrived to the Trobriand Islands in 1914 due to the outbreak of World War I, his subsequent four-year immersion in island society, and the development of participant observation methodology. The introduction emphasizes that while the game can only approximate Malinowski's complex ethnographic experience, players can gain valuable insights into both Trobriand culture and pioneering anthropological methods through this interactive engagement.

\textbf{Quizz Improvement:} We refined the quiz functionality. In addition to the standard four answer choices, users may now request up to two hints from GPT-4o and remove two incorrect options per quiz. This feature is depicted in Figure 8 (See appendix \ref{additionalscreenshots}). 

We conducted follow-up sessions with participants P1, P3, and P10 to assess the effectiveness of the new features. The anthropologist (P3) confirmed that the revised contextual introduction text was appropriate for educational use. The two general users (P1 and P10) highlighted the value of the new functionalities, noting that the ability to look up local idioms directly from the book significantly enhanced their comprehension. They also described the additional quiz features as “making the game more fun”. Both participants praised the interactive book consultation functionality; for example, P10 remarked, “It’s a great idea to have users chat with the book—it allows for more immersive study”. Overall, participants emphasized that these features shifted their engagement even more from passive reading toward active scholarly inquiry.

\section{Limitations}
\label{limitations}
Our work consisted of three main phases: building the prototype, evaluating it, and expanding on it through iterative design suggestions. All phases revealed important limitations that highlight areas for future improvement.

The most significant limitation concerns the exclusion of indigenous voices from the production process. This absence creates a substantial risk of reproducing colonial perspectives by developing content about the Trobriand Islands community without their direct involvement. Future iterations should meaningfully engage members of the Trobriand Islands or recognized experts with deep fieldwork experience in the region, such as anthropologists who have undertaken sustained ethnographic work in the area. While such collaboration would require considerably greater development resources and extended timelines, it represents an essential step toward creating more ethically grounded educational tools.

Closely related to this participation issue, the current gameplay maintains an exclusively anthropologist-centered viewpoint, reinforcing the colonial gaze inherent in early ethnographic methodology. This design choice perpetuates the problematic dynamic of observing rather than engaging with indigenous knowledge systems. The game's focus on Malinowski's perspective, while educationally valuable for understanding the history of anthropological method, risks normalizing colonial approaches to cultural study. To address this recognized limitation, we plan to accompany the game with comprehensive documentation that explicitly acknowledges these concerns and directs users to contemporary decolonial scholarship. This will include recent critical analyses of Malinowski's work, such as \citet{steinmuller2024archaeology} examination of Malinowski's economic frameworks, which provides important counterbalancing perspectives to the colonial narrative.

Additionally, while our evaluation approach successfully generated both quantitative and qualitative insights about the prototype's effectiveness, the user sample presented significant constraints. Participants were relatively few in number and drawn predominantly from a highly educated demographic, consisting primarily of master's and doctoral degree holders. This narrow sample limits our understanding of how the system might perform across diverse educational contexts and learning backgrounds. The homogeneous nature of our participant pool raises questions about the prototype's accessibility and effectiveness for undergraduate students or learners from different disciplinary backgrounds. Future research should incorporate broader participant pools, including studies within undergraduate anthropological education settings, to develop a more comprehensive understanding of the prototype's educational impact and to ensure that the tool serves diverse learning needs effectively.

During the expansion phase, where we gathered design suggestions for iterative improvement, a significant constraint emerged regarding the scope of implementation. While we collected comprehensive feedback and identified numerous potential enhancements, time and resource limitations restricted us to implementing only four of the proposed design suggestions. More sophisticated features, such as a comprehensive content moderation system or using different LLMs, remained beyond the scope of this iteration. These advanced capabilities would require substantially more development time and technical infrastructure to implement effectively.

\section{Ethical Considerations}
\label{Ethics}
This research centers on societal interests with a primary focus on public good. Our ultimate aim is to cultivate player interest and curiosity about anthropological texts—including monographs, articles, and field reports—thereby potentially expanding social anthropology's readership. However, several ethical dilemmas warrant careful consideration.

Firstly, concerns exist regarding the potential displacement of human creators—game designers, artists, and programmers—through our LLM-based approach to game development. This concern mirrors broader societal anxieties about AI-driven unemployment, particularly in creative sectors \cite{erickson2024ai}. However, this apprehension appears largely unfounded in the specific context of anthropological games, as gaming has historically had minimal presence within social anthropology \cite{harper2018games}, with few anthropologists venturing into game production. Rather than displacing an existing economic ecosystem, our approach may instead catalyze the emergence of specialized developers or even game development studios focused on developing AI-native anthrogames, potentially creating new employment opportunities in this nascent field.

Secondly, the environmental impact of scaled implementation merits serious attention. \citet{samuel2021design} highlight the substantial energy demands associated with AI-driven gameplay, citing "Deepmind's AlphaStar Final" as exemplifying the enormous power requirements for training such systems. The environmental footprint remains a significant consideration for any deployment strategy. Future development must prioritize energy-efficient algorithms and sustainable computing practices to mitigate these environmental concerns that come along with AI-native anthrogames.

Thirdly, as with other generative AI applications in educational contexts, there exists the risk of fostering student overdependence on AI technologies for academic comprehension \cite{pratiwi2025transforming}. This potential dependency could undermine the development of critical thinking skills and independent analysis capabilities that are fundamental to anthropological inquiry \cite{zhai2024effects}. Longitudinal studies examining how students interact with AI-native educational games will be essential to understand and address this concern effectively.
By acknowledging these ethical considerations transparently, we position our research to contribute meaningfully to both anthropological education and responsible AI development in educational contexts.

\section{Conclusion}
\label{conclusion}
This study introduced Malinowski’s Lens, an AI-based educational game for anthropology. The system integrates retrieval-augmented generation with GPT-4o for narrative guidance and content scaffolding, and DALL·E 3 for text-to-image generation, yielding interfaces that are coherent and adaptable while supporting dynamic content creation.

The main HCI contribution is responsive ethnographic simulation: user actions elicit contextually relevant cultural learning moments. Using a two-phase interaction model—fieldwork exploration and academic defense — the system shows how exploratory learning can be paired with structured assessment while preserving user agency in AI-mediated environments.


Our empirical evaluation provides robust evidence for the effectiveness of AI-native educational interfaces. Study 1 with 10 non-specialist participants demonstrated strong learning outcomes (quiz scores ranged 6-9/10, average 7.5) and excellent usability (SUS: 83/100), with participants expressing increased interest in reading the original text. Study 2 with 4 expert anthropologists confirmed pedagogical value, with one senior anthropologist discovering "new aspects" of Malinowski's work through our interface. Expert endorsement for classroom integration validates the system's practical utility as an active learning supplement. Building on participant feedback, we also implemented four of the proposed improvements to the game. These enhancements were tested with a subset of the original 14 playtesters, who responded positively and approved these changes.

Our ethical interface design approach—representing indigenous peoples as silhouettes while depicting Malinowski in detail, demonstrates how design decisions can address both technical limitations in AI-generated cultural representation and prompt critical reflection on anthropological representation practices. This work contributes to broader discussions in HCI about responsible AI interface design and ethical considerations in culturally sensitive interactive systems.

The technical framework we present offers a replicable model for transforming dense academic texts into engaging interactive experiences. By demonstrating successful integration of RAG-based narrative generation with consistent visual world-building, this work establishes new possibilities for AI-native educational interfaces across disciplines. The system's success in conveying complex anthropological concepts while fostering disciplinary curiosity suggests broad applicability for digital humanities and educational technology. 

Finally, this research advances our understanding of how generative AI can create novel forms of human-computer interaction in educational contexts. As AI technologies continue to evolve, the principles and techniques demonstrated in Malinowski's Lens provide a foundation for developing more sophisticated AI-mediated learning environments.

\section{Future Work}
\label{futurework}
Our future research agenda encompasses two key directions: 

Firstly, we will refine 'Malinowski's Lens' through iterative user studies with two distinct target groups: anthropology students in university programs and members of the communities described in the anthropological monograph itself. These studies will inform evidence-based improvements to the game's design and educational effectiveness.


Secondly, we aim to develop a no-code platform that lowers the entry barrier for the creation of educational games from anthropological texts. This platform will enable educators, museum curators and researchers without programming expertise to transform scholarly works into interactive learning experiences, potentially expanding the reach and impact of anthropological knowledge through gamification.

\bibliographystyle{ACM-Reference-Format}

\bibliography{sample-base}

\clearpage

\section*{Appendix}
\appendix
\label{appendix}

\section{Prompt Example for Image Generation}
\label{imagegeneratonprompt}
This section presents an example text-to-image generation prompt for DALL-E 3 in JSON format, designed to create pixel-art style ethnographic scenes from Malinowski's work.
\begin{verbatim}
{
"content": "You are an assistant that writes DALL·E-compatible image
prompts in a single line. You are a pixel-art renderer in the style
of 1991–94 320×240 adventure games.
STYLE RULES:
Canvas 320×240, ~32 colours max, 1-px dark outlines.

Use classic dithering for mid-tones, no smooth gradients or bloom.

Islanders must appear as dark silhouettes; Malinowski in full detail.

No text, UI, or modern effects.

OUTPUT: base-64 PNG, pixel-perfect 1:1.
Always begin each prompt with 'pixel-art:' so the style is consistent.
All scenes, characters, and artifacts must be grounded strictly in the
ethnographic work 'Argonauts of the Western Pacific' by Bronislaw
Malinowski. Depict only realistic cultural practices, environments,
and artifacts from this source. Do not include magical, fantastical,
or fictional elements.
Inputs you will receive every turn
    day_number – integer

    narrative_excerpt – 40-60 word summary of the current scene

    new_artifact_names[] – list of artifact ids just obtained 
    (may be empty)
Return one prompt for the scene plus (if list non-empty) one prompt
per artifact—each on its own line."
}
\end{verbatim}

\clearpage
\section{Post-Play Qualitative Interview Framework}
\label{survey}
\begin{table}[htbp]
  \caption{Post-Play Qualitative Interview Framework}
  \label{tab:interview_guide}
  \small
  \begin{tabular}{@{}p{0.23\textwidth} p{0.74\textwidth}@{}}
    \toprule
    \textbf{Theme} & \textbf{Interview questions} \\
    \midrule
    Anthropological curiosity &
      \begin{itemize}
        \item Which aspects of Trobriand culture sparked your curiosity to learn more?
        \item Did the game motivate you to read Malinowski’s original text? If so, why?
        \item How, if at all, did the game influence your interest in exploring other anthropological studies?
      \end{itemize} \\
    \addlinespace
    Game design &
      \begin{itemize}
        \item Which visual design choices best supported your understanding of Trobriand culture?
        \item Did the game effectively convey its educational purpose?
        \item Which aspects of the game stood out?
        \item What improvements would you suggest for the game?
      \end{itemize} \\
    \addlinespace
    General feedback &
      Any other thoughts you would like to share? \\
    \bottomrule
  \end{tabular}
\end{table}
\clearpage

\section{General User Individual Quiz Scores}
\label{individualquizzscores}
\begin{table}[htbp]
  \caption{General User Individual Quiz Scores}
 
  \begin{tabular}{l r}
    \toprule
    \textbf{Participant ID} & \textbf{Score} \\
    \midrule
    \quad 1  & 7  \\
    \quad 2 & 9 \\
    \quad 4  & 8  \\
    \quad 7 & 8 \\
    \quad 8  & 8  \\
    \quad 10 & 7 \\
    \quad 11  & 6  \\
    \quad 12 & 8 \\
    \quad 13  & 7  \\
    \quad 14 & 7 \\
     \midrule
    \textbf{Mean (Total)} & \textbf{0.75} \\
    \textbf{Median (Total)} & \textbf{0.75}\\
    \textbf{Std Var (Total)} & \textbf{0.81} \\
    \bottomrule
  \end{tabular}
\end{table}

\section{Additional Screenshots from Malinowski's Lens (Iterated Version)}
\label{additionalscreenshots}
\begin{figure}[htbp]
  \centering
  \includegraphics[width=0.75\columnwidth]{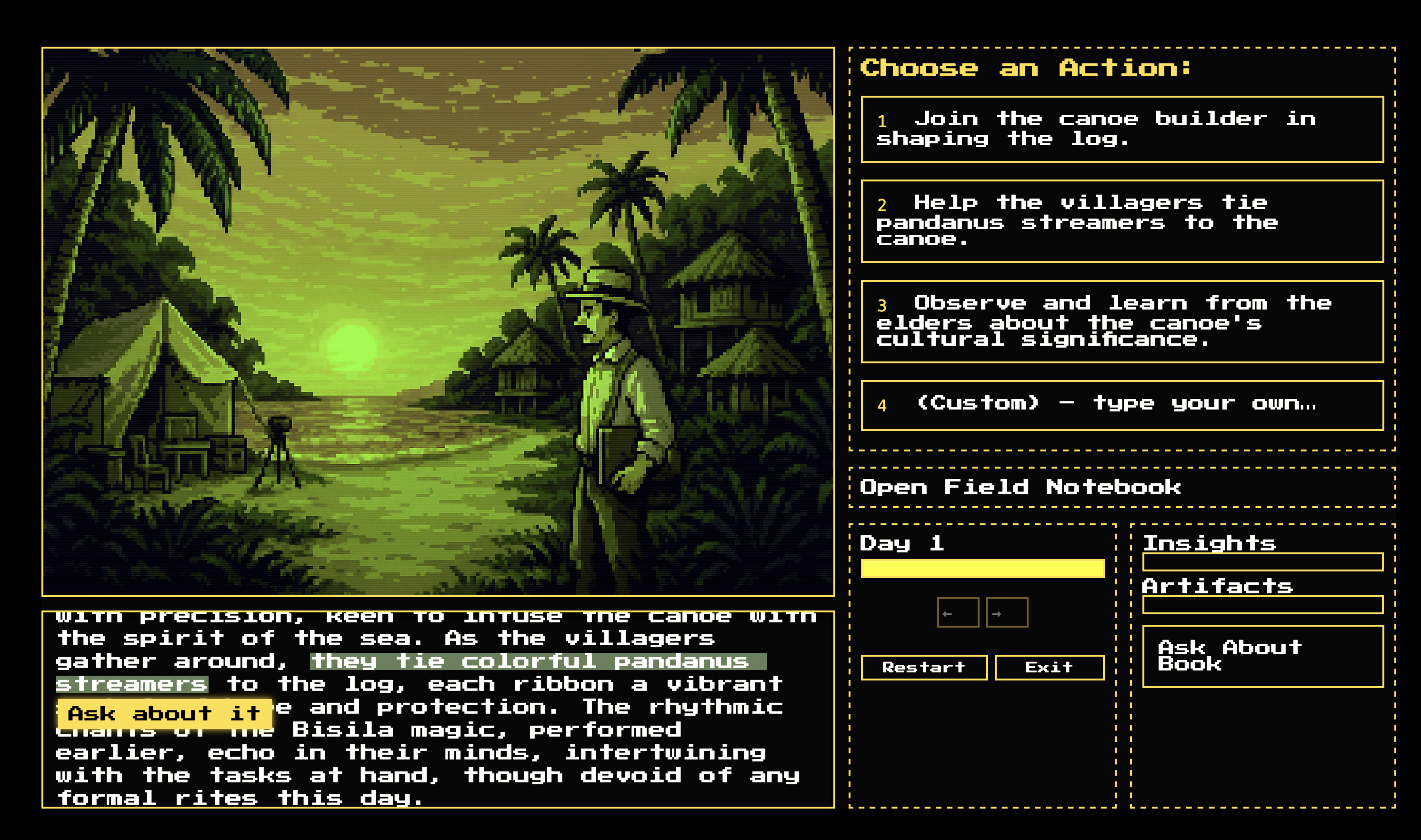}
  \caption{Updated Malinowski's Lens interface screenshot featuring the new 'Ask about it' option that lets users query unfamiliar local idioms within the narrative section shown at bottom left.}
  \Description{Screenshot illustrating the interface used to request explanations about in-game scenes.}
  \label{fig:feature1}
\end{figure}

\begin{figure}[htbp]
  \centering
  \includegraphics[width=0.75\columnwidth]{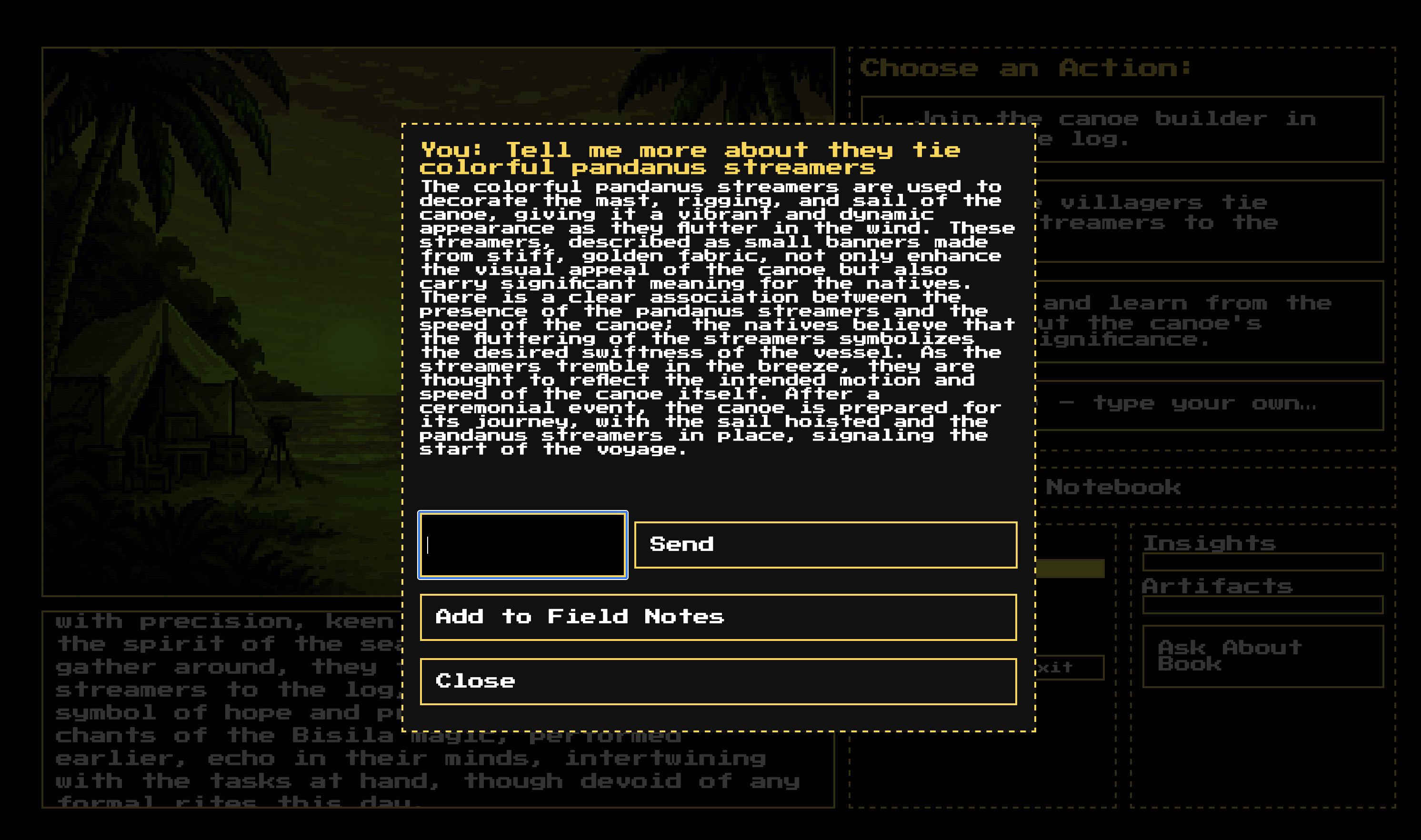}
  \caption{Follow-up screenshot showing the detailed explanation that appears after clicking 'Ask about it' on an unknown local idiom}
  \Description{Second screenshot showing a follow-up scene explanation and response.}
  \label{fig:feature2}
\end{figure}

\begin{figure}[htbp]
  \centering
  \includegraphics[width=0.75\columnwidth]{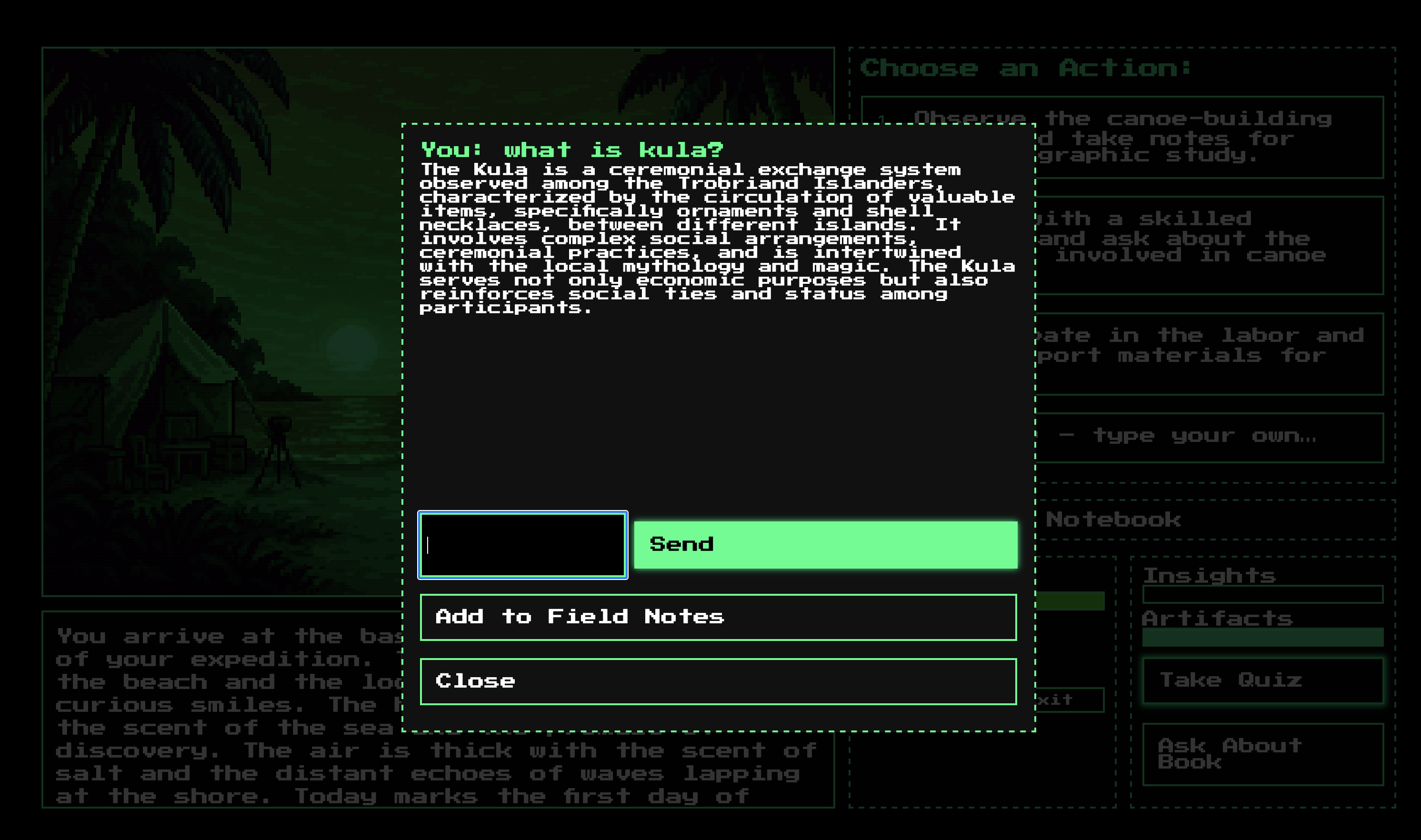}
  \caption{Interactive Book consultation feature, enabling direct inquiries about the source text to support immersive literary study during gameplay.}
  \Description{Screenshot of the in-game book consultation feature used for contextual learning.}
  \label{fig:feature3}
\end{figure}

\begin{figure}[htbp]
  \centering
  \includegraphics[width=0.75\columnwidth]{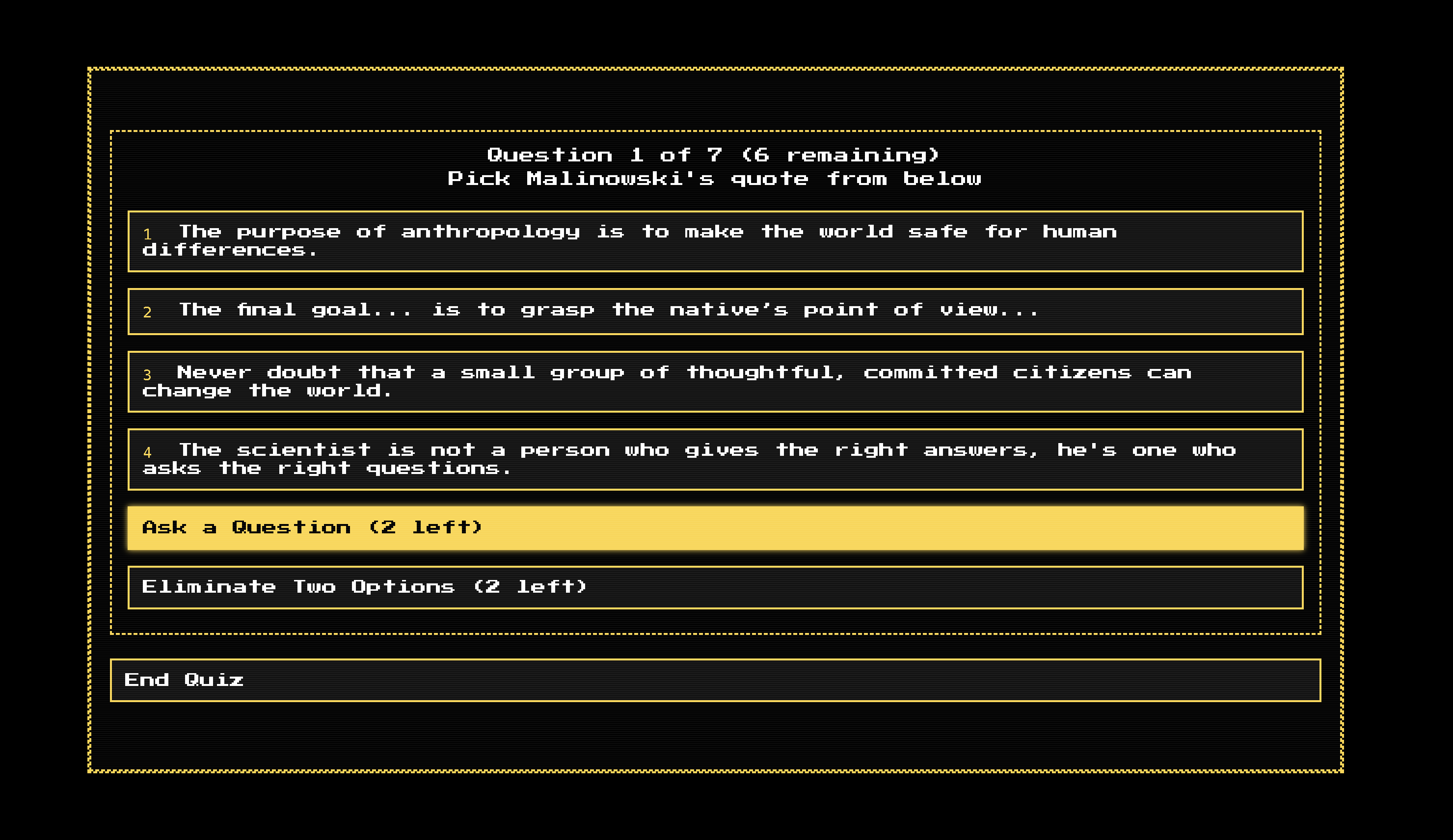}
  \caption{Enhanced quiz interface with features to request a hint and eliminate two answer options}
  \Description{Improved Quizz}
  \label{fig:feature3}
\end{figure}

\end{document}